\documentclass[conference]{IEEEtran}
\IEEEoverridecommandlockouts
\usepackage{cite}
\usepackage{amsmath,amssymb,amsfonts}
\usepackage{algorithmic}
\usepackage{graphicx}
\usepackage{textcomp}
\usepackage{xcolor}
\usepackage{eurosym}

\def\BibTeX{{\rm B\kern-.05em{\sc i\kern-.025em b}\kern-.08em
    T\kern-.1667em\lower.7ex\hbox{E}\kern-.125emX}}
\begin{document}

\title{Electric Vehicle Fleet Relocation Management for Sharing Systems based on Incentive Mechanism\\
\thanks{This work is a part of the ELVITEN and NeMo projects. ELVITEN has received funding from the European Union’s Horizon 2020 research and innovation programme under Grant Agreement no 769926. NeMo has received funding from the European Unions Horizon 2020 research and innovation programme under Grant Agreement no 713794. Content reflects only the authors’ view and European Commission is not responsible for any use that may be made of the information it contains.}
}

\author{\IEEEauthorblockN{Maria Pia Fanti, Agostino Marcello Mangini, Michele Roccotelli}
\IEEEauthorblockA{\textit{Department of Electrical and Information} \\
\textit{Engineering} \\
\textit{Polytechnic University of Bari}\\
Bari, Italy \\
(mariapia.fanti,agostinomarcello.mangini,michele.roccotelli)@poliba.it}
\and
\IEEEauthorblockN{Bartolomeo Silvestri, Salvatore Digiesi}
\IEEEauthorblockA{\textit{Department of Mechanics, Mathematics} \\
\textit{and Management} \\
\textit{Polytechnic University of Bari}\\
Bari, Italy \\
(bartolomeo.silvestri,salvatore.digiesi)@poliba.it}
}


\maketitle

\begin{abstract}
This paper deals with the electric vehicle fleet relocation management in a sharing system. The mobility sharing systems efficiency depends on the vehicles relocation task that strongly affect the company operating cost, and consequently the service price for users. The proposed approach aims at minimizing the cost of vehicles relocation for a sharing company by involving users through an innovative incentive scheme.
The idea is to request users of the sharing service to relocate the EVs, e.g. through an IT application, incentivizing them by free-of-charge travels and rewards. The proposed incentive mechanism is based on the application of different levels of incentive proposal. In addition, in case of user unavailability, the vehicle relocation is guaranteed by the company staff. To this aim, a first ILP is formalized to manage the relocation task by the company staff. Moreover, a second ILP allows the company to involve users in the relocation process by the proposed incentive mechanism. Finally, a case study is presented to show the application of the proposed methodology on the relocation of electric cars and light electric vehicles.
\end{abstract}

\begin{IEEEkeywords}
Electric Vehicle, Vehicle Relocation, Sharing System, User Incentive
\end{IEEEkeywords}

\section{Introduction}
Nowadays, the increasing need for improving the air quality in urban areas motivates the introduction of green transport means and of new
mobility solutions able to reduce the pollutant emissions and the traffic congestion. 
Car sharing (CS) is today one of the most popular solution, based on the possibility for a car to be used as a public transport means
by individual user that can autonomously rent the vehicle according to his/her needs, usually for a short time period \cite{Clemente}. One of the main issue of CS is to maintain a high level of the service efficiency despite the dynamic reconfiguration of the system that cyclically occurs during its operation. Indeed, periodic relocation of vehicles between stations becomes necessary to ensure that there are sufficient vehicles that are spread geographically across the stations to serve user demands \cite{DSS1}. Two different rental strategies can be distinguished in the car sharing service \cite{Clemente}: i) the one-way rental; ii) the two-way rental. The majority of the researchers focus on the first strategy, where users or operators are allowed to pick up and return the rented vehicle in different parking areas \cite{Clemente},[2],[4]-[19]. In this case, the distribution of the vehicles can become unbalanced during the day due to the variable demand \cite{Clemente}. In the second strategy, the car sharing users or operators have to return the car to the pick-up station \cite{two-way},[7]. In this way, the number of vehicle per station can remain constant but the user have no flexibility in their travels. For all these reasons, it is essential to solve the vehicle relocation problem within a car sharing system. With this aim, different strategies and solutions have been proposed by researchers and practitioners in the last years.
There are user-based approaches that aim at incentivizing the users to participate in the relocation activities, balancing the vehicles among the parking areas \cite{Clemente},[4],[7]-[13],[16]. In this case, vehicle relocation is done by users that travel, mainly through incentives in the form of trip cost discount. On the other hand, there are operator-based approaches, where system staff are asked to relocate the vehicles when it is necessary [5]-[7], [13]-[19]. In this case, additional trips without customers are needed by the operators to relocate the vehicles. The first approach should be preferred because allows for the reduction of traffic congestion, environmental pollution and costs. 
A further classification of car sharing relocation strategies can be done based on  relocation time: \textit{off-line} [15]-[19] or \textit{real-time} \cite{Clemente}-[14]. 

This paper proposes a one-way, operator-based and user-based, off-line relocation approach.
More recently, in the proposed free-floating approaches it is possible to pick-up and return the vehicle to almost any parking spot within a defined operating area [5], [10], [19]. 
It is remarked that most contributions in related literature focus on the relocation problem of conventional Internal Combustion Engine vehicles. On the contrary, few papers consider the introduction of electric vehicles (EVs) in sharing systems, which brings additional constraints to the relocation problem, mainly related to the charge requirements [5], [7], [14], [17]. In this context, all the analyzed papers propose operator-based approaches and, in most of cases, no incentives are given for the EVs relocation. 

Differently from the analyzed works, in this paper, we aim at solving the EVs relocation problem in sharing systems by involving customers through an innovative incentive mechanism based on free-of-charge travels and rewards. The proposed incentive mechanism applies different levels of incentives and in case of user unavailability, the vehicle relocation is guaranteed by the company staff. To this aim, a first ILP is formalized to manage the relocation task by the company staff. Moreover, a second ILP allows the company to involve users in the relocation process by the proposed incentive mechanism. Finally, a case study shows the application of the proposed methodology on the relocation of electric cars and Electric Light Vehicles (ELVs).

The rest of the paper is organized as follows. Section II presents the Electric Vehicle Relocation Problem (EVRP) and Section III propose an optimization approach to solve it. Furthermore, Section IV presents a case study and potential benefits of applying the proposed optimal relocation approach to EVs and ELVs fleet are shown. Finally, Section V summarizes conclusions and future perspectives.
 
\section{Electric Vehicle Relocation Problem}
The diffusion of shared mobility services is limited due to the price for the end users, especially for some categories such as young people, the limited users information and the lack of experience in using such as new mobility systems and vehicles. In order to reduce the cost for the users of sharing services, and at the same time guaranteeing the economic sustainability of the company, the following main strategies can be applied [2]: 
\begin{itemize}
\item reduction of operating costs;
\item increase of the daily usage rate of vehicles.
\end{itemize}
In this context, vehicle relocation is important to increase the quality level of the sharing service, making vehicles available in different stations/areas at demand peaks daytime. However, the vehicle relocation is an expensive task for the company that needs to be reduced. Indeed, the reduction of this cost will increase the company profitability and will allow to reduce the price of the sharing service for the customers. 

The proposed approach aims not only at reducing the cost of relocation for a mobility sharing company, but also at attracting and retaining customers through an incentive mechanism. The customers that use the mobility sharing service for their trips can also become active part of the relocation process with direct benefits for both the sharing company and the users.

The idea is to request users to move a vehicle from one station to another being rewarded. In particular, the user that accept to participate in vehicle relocation will travel for free and will receive trip credits to be used in the sharing service. It is remarked that, the cost of providing trip credits to customers is lower than the cost of the relocation service for the sharing company. In addition, only EVs with  battery autonomy that allows to accomplish the relocation trip can be relocated. Nowadays, mobility sharing companies provide their services through an IT application for smartphone.

\begin{figure}[thpb]
      \centering
      \includegraphics[scale=0.35]{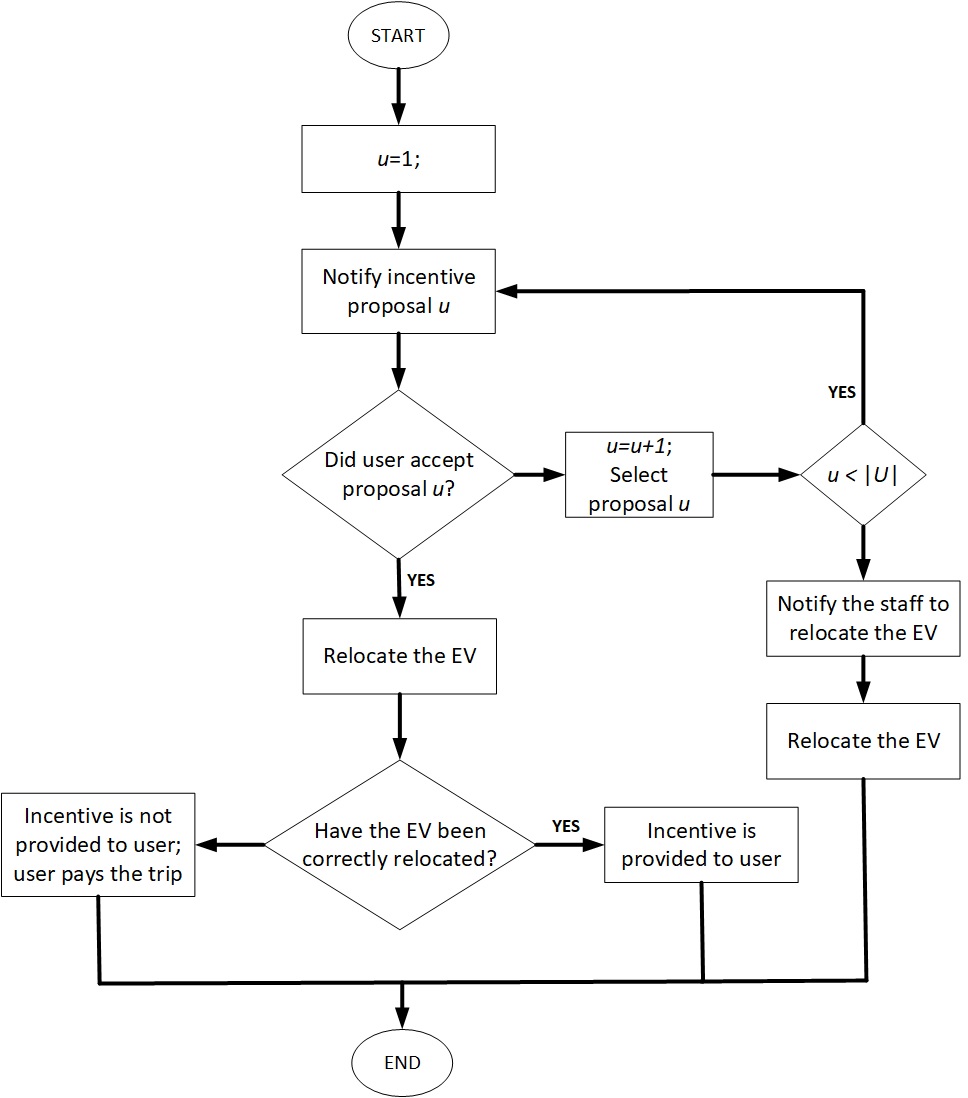}
      \caption{The incentive-based relocation  process}.
      \label{relo_pro}
\end{figure}

The proposed user incentivization approach can be provided through an IT application and can be described by the following procedure (see Fig. \ref{relo_pro}). First, an incentive bid is notified to all users who want to participate in the relocation activities. At this stage, two cases can occur: i) at least one user per vehicle to be relocated accepts the incentive proposal, pick-up the vehicle at the departure station and go to the destination station and receive the reward at the end of the relocation activity (the reward is loaded on his/her sharing service account); ii) no user accepts the incentive proposal and a new notification is broadcast to all users with an increased incentive bid. Note that step ii) can be repeated several times based on the incentive levels that the company propose (e.g. 3 levels of incentives means that step ii can be repeated at most three times). After that, if no user accepts the incentives proposals, the company will use internal staff to perform the relocation service. On the other hand, the standard service fee and no incentive will be applied to users who accept the incentive proposal but deliver the vehicle to a different station.

The proposed incentive mechanism allows reducing operational costs of vehicle relocation or at worst to keep them unchanged. 
The EV relocation incentive-based process for a sharing company is depicted in Fig. \ref{relo_pro}. Let us introduce the set of incentive proposals (levels) $\mathcal{U} \in \mathbb{Z^+}$, with cardinality $\vert \mathcal{U} \vert$. It is supposed that $u \in \mathcal{U}$, $u=1,\dots, \vert \mathcal{U} \vert-1$ are the incentive levels planned by the company to the users. Furthermore, we assume that  the last element of $\mathcal{U}$ does not represent an incentive for user but it is used to enable the relocation process by the company staff.  

In this paper, both electric cars and light electric vehicles (ELVs) are considered [20]. In particular, ELV is a small size EV with 2, 3 or 4 wheels, such as mopeds and motorbikes, quad, tricycles and quadricycles. In the proposed relocation approach, specific types of ELVs, such as e-bikes, or of EVs such as segway and electric kick scooters, can be relocated even in case of low battery autonomy but only by the company staff. Indeed, due to their size it is possible pick them up by means of a van.

\section{Optimization Approach for Electric Vehicle Relocation}
This section presents the optimization approach adopted to solve the EVRP for a sharing company by involving users with incentive proposals. In the proposed framework, the sharing system is station-based and also a restricted area can be considered as a station. The optimization model is formulated under the following assumptions: 
\begin{itemize}

\item the relocation cost for the company is a variable cost proportional to the distance between stations based on a predefined destination-source matrix and does not include the fixed cost of the company (e.g. salary, etc.);
\item the relocation process is off-line: all the EVs are assumed to be in the stations before the relocation process starts;
\item the number of daily relocation activities does not change if performed by customers or by company staff;
\item the battery autonomy of each EV is known;
\item in the user relocation process the EVs are moved from one station to another through the shortest route and only EV requested by company are picked-up.
\end{itemize}

\subsection{Electric Vehicle Relocation Problem for the Company Staff}
In this subsection the relocation problem is formulated as an integer linear programming (ILP) problem that aims at minimizing the cost of the EV relocation for the company. Given the set $\mathcal{K}\in\mathbb{Z^+}$ of EVs to be relocated, with cardinality $\vert \mathcal{K} \vert$, and the set $\mathcal{I} \in\mathbb{Z^+}$ of stations, with cardinality $\vert \mathcal{I} \vert$. On a daily base, the relocation service cost depends on the number of EVs to be relocated and from the predefined distance among the stations, which are assigned. The binary decision variables are $x_{i,j,k}$, $\forall i,j \in \mathcal{I}$, $\forall k \in \mathcal{K}$, where the departure station is labeled with $i$, the arrival station is labeled with $j$ and the EV is labeled with $k$.

The integer variables are $s_{j}$, $\forall j \in \mathcal{I}$, indicating the number of EVs in each station after the relocation process. Moreover, we define $\mathcal{NK}_{i,j} \in \mathbb{N}_0$ the set of EVs that can be relocated from station $i$ to station $j$, based on the battery autonomy, with cardinality $\vert \mathcal{NK} \vert=nk_{i,j}^{max}$.  
The known parameters of the problem are the following: $\hat{s}_{j}$ is the number of EVs in station $j$ before the relocation process; $\hat{s}_{i}$ is the number of EVs in station $i$ before the relocation process; the maximum and minimum number of EVs to be guaranteed at each station, are respectively defined as $N_{max} \in \mathbb{Z^+}$ and $N_{min} \in \mathbb{Z^+}$.

Hence, the following variables and parameters are defined: 
\begin{itemize}
\item $RC$ is the total relocation cost for the mobility sharing company [\euro /day];
\item $c$ is the cost per kilometer for the relocation task [\euro /km];
\item $D_{i,j}$ is the distance from station $i$ to station $j$ [km];
\item $x_{i,j,k}$ is the binary decision variable indicating the $k$-th EV moving from the station $i$ to station $j$ in one day [day$^{-1}$];
\item $x_{j,i,k}$ is the binary decision variable indicating the $k$-th EV moving from station $j$ to station $i$ in one day [day$^{-1}$];
\item $s_j$ is the number of EVs in station $j$ after the relocation process;
\item $\hat{s}_{j}$ is the number of EVs in station $j$ before the relocation process;
\item $v_{i,k}=1$ means that EV $k$ has sufficient battery autonomy to move from station $i$ to station $j$;
\item $a_{i,k}$ is the battery autonomy of EV $k$ in station $i$ [km];
\item $N_{max}$ is the maximum admissible number of EVs in the $j$ station; 
\item $N_{min}$ is the minimum admissible number of EVs in the $j$ station;
\item $nk_{i,j}^{max}$ is the maximum number of EVs with sufficient battery autonomy to move from station $i$ to station $j$. It is calculated as:
\end{itemize}
\begin{equation}\label{eqB}
nk_{i,j}^{max}= \sum_{k=1}^{\hat{s}_{i}} \; v_{i,k}, \; \forall i \in \mathcal{I}, \;
\end{equation}
with
\begin{equation}\nonumber
\resizebox{.4\hsize}{!}{$
v_{i,k}= \begin{cases} 1$, if $a_{i,k}-D_{i,j}\geq 0\\
0$, otherwise
$\end{cases}$}
\end{equation}

The ILP1 (\ref{ILP1})-(\ref{eq5}) is formulated as follows:

\begin{equation}\label{ILP1} 
RC=\min \sum_{i=1}^{\vert \mathcal{I} \vert} \; \sum_{j=1, i \neq j}^{\vert \mathcal{I} \vert} \; \sum_{k=1}^{\vert \mathcal{K} \vert} \; c \cdot D_{i,j} \cdot x_{i,j,k}
\end{equation}

 s.t.\\
\begin{equation}\label{eq1}
s_{j}= \hat{s}_{j} + \sum_{i=1}^{\vert \mathcal{I} \vert} \; \sum_{k=1}^{\vert \mathcal{K} \vert} \; x_{i,j,k} \ - \sum_{i=1}^{\vert \mathcal{I} \vert} \; \sum_{k=1}^{\vert \mathcal{K} \vert} \;  x_{j,i,k}, \; \forall j \in \mathcal{I}
\end{equation}

\begin{equation}\label{eq2}
s_{j} \leq N_{max}, \; \forall j \in \mathcal{I}
\end{equation}

\begin{equation}\label{eq3}
s_{j}\geq N_{min}, \; \forall j \in \mathcal{I} 
\end{equation}

\begin{equation}\label{eq4}
\sum_{k=1}^{\vert K \vert} \; x_{i,j,k} \leq nk_{i,j}^{max}, \; \; \forall i,j \in \mathcal{I}, i \neq j \;
\end{equation}

\begin{equation}\label{eq5}
x_{j,i,k} \in \lbrace0,1\rbrace \;
\end{equation}

$i,j \in \mathcal{I}$, 

$k \in \mathcal{K}$

Variables number:
$\vert \mathcal{I} \vert \times(\vert \mathcal{I} \vert \times \vert \mathcal{K} \vert + 1)$

Constraints: $\vert \mathcal{I} \vert \times(\vert \mathcal{I} \vert + 2)$ 
\\

The constraints (\ref{eq4}) can lead to an unfeasible solution because constraints (\ref{eq2}) and (\ref{eq3}) can not be verified if there are no EVs for relocation due to low battery capacity.
In real cases, constraints (\ref{eq4}) can be relaxed, if the company performs relocation of ELVs by means of vans. Moreover, if the company does not employ van, then it is possible to reach a feasible ILP1 solution by properly decreasing $N_{min}$ and/or increasing $N_{max}$.

\subsection{Electric Vehicle Relocation Problem Involving Users by Incentive Mechanism}

The introduction of users in the relocation process requires a new formulation of the optimization problem (\ref{ILP1})-(\ref{eq5}). Let us recall the set of incentive proposal levels as defined in Section II. The number $N_c \in \mathbb{N}$ of users available in the relocation process is defined. The user acceptance based on the proposed incentive levels is also considered. In particular, a user acceptance rate $r_u \in \mathbb{R^+}$ is introduced for each level of incentive. The number  of relocatable EVs is given by $evr_u=N_c*r_u$, $u=1,\dots, \vert \mathcal{U} \vert-1$. It is remarked that, in the proposed approach the relocation task can be performed both by customers and by company staff in order to reduce the relocation cost.
Moreover, only EVs with necessary state of charge (SoC) to accomplish the relocation trip can be used. 

Now, the following variables and parameters are defined:

\begin{itemize}
\item $RCI$ is the total relocation cost for the mobility sharing company inclunding the incentive for users [\euro /day];
\item $in_{u}$ is the incentive rate for the incentive level $u$;
\item $x_{i,j,k,u}$ is the binary decision variable indicating the $k$-th EV moving from station $i$ to station $j$ under incentive level $u$ in one day [day$^{-1}$];
\item $x_{j,i,k,u}$ is the binary decision variable indicating the $k$-th EV moving from station $i$ to station $j$ under incentive level $u$ in one day [day$^{-1}$];
\item $nk_{i,j,u}^{max}$ is the maximum number of EVs with sufficient battery autonomy to move from station $i$ to station $j$ under incentive level $u$. It is calculated as:
\end{itemize}
\begin{equation}\label{eqB}
nk_{i,j,u}^{max}= \sum_{k=1}^{\hat{s}_{i}} \sum_{u=1}^{\vert \mathcal{U} \vert} \; \; v_{i,k,u}, \; \forall i \in \mathcal{I}, \;
\end{equation}
with \; \begin{equation}\nonumber
\resizebox{.4\hsize}{!}{$
v_{i,k,u}= \begin{cases} 1$, if $a_{i,k}-D_{i,j}\geq 0\\
0$, otherwise.
$\end{cases}$}
\end{equation}

The second optimization problem is formulated as the following ILP2 (\ref{ILP2})-(\ref{eq11}):

\begin{equation}\label{ILP2}
RCI=\min \sum_{i=1}^{\vert \mathcal{I} \vert} \; \sum_{j=1, i \neq j}^{\vert \mathcal{I} \vert} \; \sum_{k=1}^{\vert \mathcal{K} \vert} \; \sum_{u=1}^{\vert \mathcal{U} \vert} \; (in_{u} \cdot c \cdot D_{i,j} \cdot x_{i,j,k,u})
\end{equation}

s.t.\\
\begin{equation}\label{eq6}
s_{j}= \hat{s}_{j} + \sum_{i=1}^{\vert \mathcal{I} \vert} \; \sum_{k=1}^{\vert \mathcal{K} \vert} \; \sum_{u=1}^{\vert \mathcal{U} \vert} \; x_{i,j,k,u} \ - \sum_{i=1}^{\vert \mathcal{I} \vert} \; \sum_{k=1}^{\vert \mathcal{K} \vert} \; \sum_{u=1}^{\vert \mathcal{U} \vert} \; x_{j,i,k,u} \ \forall j \in \mathcal{I}
\end{equation}

\begin{equation}\label{eq7}
s_{j}\leq N_{max}, \; \forall j \in \mathcal{I}
\end{equation}

\begin{equation}\label{eq8}
s_{j}\geq N_{min}, \; \forall j \in \mathcal{I}
\end{equation}

\begin{equation}\label{eq9}
\sum_{k=1}^{\vert \mathcal{K} \vert} \; x_{i,j,k,u} \leq nk_{i,j,u}^{max}, \; \; \forall i,j \in \mathcal{I}, i \neq j, \; \; \forall u \in \mathcal{U}
\end{equation}

\begin{equation}\label{eq10}
\sum_{i=1}^{\vert \mathcal{I} \vert} \; \sum_{j=1}^{\vert \mathcal{I} \vert} \; \sum_{k=1}^{\vert \mathcal{K} \vert} \; x_{i,j,k,u} \leq evr_{u}, \; \forall u \in \mathcal{U} 
\end{equation}

\begin{equation}\label{eq11}
x_{j,i,k,u} \in \lbrace0,1\rbrace \;
\end{equation}

$i,j \in \mathcal{I}$

$k \in \mathcal{K}$

$u \in \mathcal{U}$

Variables number:
$\vert \mathcal{I} \vert \times(\vert \mathcal{I} \vert \times \vert \mathcal{K} \vert \times \vert \mathcal{U} \vert \ + 1)$

Constraints: $\vert \mathcal{I} \vert \times(\vert \mathcal{I} \vert + 2) + 2 \times \vert \mathcal{U} \vert$
\\

The same considerations done for constraints (\ref{eq4}) in ILP1, can be applied to constraints (\ref{eq9}) of ILP2, if the relocation task is performed by the company staff. On the contrary, constraints (\ref{eq9}) can not be relaxed for ILP2 if the relocation task is performed by customers. 

It is expected that the relocation cost obtained by solving ILP2 is less than or at least equal (in case no user is available for relocation) to the relocation cost obtained by solving ILP1. 

The proposed optimization approach can be applied to real EVRPs for station-based sharing systems that make use of any categories of EVs (e.g. electric car sharing, electric scooter sharing, electric bike sharing, and others).

\section{Case Study}
In this section, we present a case study to show the benefits of the proposed methodology for solving the EVRP in sharing systems.

The formulated ILP problems are solved by a standard solver, i.e. MatLab (LinProg), on a Intel-Core i5, 2,7 Ghz CPU with 8 GB RAM. All the performed tests are solved in few seconds.

Two scenarios are proposed to solve ILP1 and ILP2 for two different kinds of vehicles: 1) electric cars; 2) electric bikes. 
In both Scenarios the following parameters are set: $\vert \mathcal{I} \vert=60$ (number of vehicles); $\vert \mathcal{K} \vert=6$ (number of stations); $N_{max}=20$ (20 parking slots and charging points for each station); $N_{min}=5$; $N_c=200$; $\hat{s}_j$, $\forall j \in I$ are defined in Table \ref{init_value}; $c = 1$ [\euro /km]; $D_{i,j}$, $\forall j \in \mathcal{I}$, $\forall i \in \mathcal{I}$ are defined in a destination-source matrix and reported in Table \ref{cost}; $a_{i,k}$, $\forall i \in \mathcal{I}$, $\forall k \in \mathcal{K}$ are shown in Table \ref{Autonomy}.

\begin{table}
\centering
\caption{Initial number of EVs per station}
\scalebox{1.1}{
\begin{tabular}{c|c}
\hline 
& EVs initial number ($\hat{s}$) \\ 
\hline 
Station 1 & 18 \\ 
 
Station 2 & 14 \\ 
 
Station 3 & 2 \\ 
 
Station 4 & 4 \\ 

Station 5 & 3 \\ 
 
Station 6 & 19 \\ 
\hline  
\end{tabular}}
\label{init_value}
\end{table}

\begin{table}
\centering
\caption{Distance between stations in km}
\scalebox{1.1}{
\begin{tabular}{c|cccccc}
\hline 
& S1 & S2 & S3 & S4 & S5 & S6 \\ 
\hline
S1 & - & 5 & 8 & 4 & 4 & 9 \\ 
 
S2 & 5 & - & 6 & 7 & 10 & 10 \\ 

S3 & 8 & 6 & - & 3 & 8 & 5 \\ 
 
S4 & 4 & 7 & 3 & - & 3 & 2 \\ 

S5 & 4 & 10 & 8 & 3 & - & 4 \\ 
 
S6 & 9 & 10 & 5 & 2 & 4 & - \\ 
\hline 
\end{tabular}}
\label{cost}
\end{table}

By comparing Table \ref{cost} and Table \ref{Autonomy} unfeasible relocation trips can be derived.

\begin{table}
\centering
\caption{EVs battery autonomy}
\scalebox{0.8}{
\begin{tabular}{ccc|ccc|ccc}
\hline
Station & EV & Autonomy & Station & EV & Autonomy & Station & EV & Autonomy \\ 
 &  & km &  &  & km &  &  & km \\ 
\hline 
1 & 1 & 4 & 2 & 21 & 5 & 5 & 41 & 4 \\ 

1 & 2 & 1 & 2 & 22 & 4 & 6 & 42 & 1 \\ 

1 & 3 & 2 & 2 & 23 & 5 & 6 & 43 & 1 \\ 

1 & 4 & 1 & 2 & 24 & 4 & 6 & 44 & 1 \\ 

1 & 5 & 1 & 2 & 25 & 3 & 6 & 45 & 1 \\ 
 
1 & 6 & 2 & 2 & 26 & 2 & 6 & 46 & 1 \\ 

1 & 7 & 3 & 2 & 27 & 4 & 6 & 47 & 1 \\ 
 
1 & 8 & 2 & 2 & 28 & 3 & 6 & 48 & 1 \\ 
 
1 & 9 & 3 & 2 & 29 & 10 & 6 & 49 & 1 \\ 

1 & 10 & 2 & 2 & 30 & 10 & 6 & 50 & 1 \\ 
 
1 & 11 & 3 & 2 & 31 & 10 & 6 & 51 & 1 \\ 

1 & 12 & 3 & 2 & 32 & 10 & 6 & 52 & 1 \\ 

1 & 13 & 2 & 3 & 33 & 2 & 6 & 53 & 1 \\ 
 
1 & 14 & 2 & 3 & 34 & 2 & 6 & 54 & 1 \\ 

1 & 15 & 3 & 4 & 35 & 3 & 6 & 55 & 1 \\ 
 
1 & 16 & 3 & 4 & 36 & 3 & 6 & 56 & 1 \\ 

1 & 17 & 3 & 4 & 37 & 3 & 6 & 57 & 1 \\ 

1 & 18 & 2 & 4 & 38 & 3 & 6 & 58 & 1 \\ 

2 & 19 & 5 & 5 & 39 & 4 & 6 & 59 & 1 \\ 

2 & 20 & 5 & 5 & 40 & 4 & 6 & 60 & 2 \\ 
\hline 
\end{tabular}}
\label{Autonomy}
\end{table}

The proposed incentive scheme is based on three levels and the incentive rates $in_{u}$ are set as follows: i) reward equal to 50\% of the company relocation service cost; ii) reward equal to 70\% of the company relocation service cost and, iii) reward equal to 90\% of the company relocation service cost. Therefore, the cardinality of set $\mathcal{U}$ is $\vert \mathcal{U} \vert=4$. In addition, the user acceptance rate $r_u$, $u=1,2,3$ are shown in Table \ref{Acc}. 

\begin{table}
\centering
\caption{User acceptance rate and relocatable EVs}
\scalebox{1.1}{
\begin{tabular}{c|cc}
\hline
& Acceptance rate & Relocatable EVs \\ 
& of users & ($evr_{u}$) \\ 
\hline 
Incentive level 1 & 0,5\% & 1 \\ 
 
Incentive level 2 & 0,5\% & 1 \\ 
 
Incentive level 3 & 1\% & 2 \\ 
\hline 
\end{tabular}}
\label{Acc}
\end{table}


\subsection{SCENARIO 1: Electric Car Sharing System}
In the first scenario, only electric cars must be relocated. By solving ILP1, the relocation cost for the sharing company by only using staff is $RC=34$ \euro /day. On the contrary, by involving users in the relocation process and solving ILP2 the relocation cost is $RCI=26$ \euro /day, with a corresponding reward $\sum_{i=1}^{6} \; \sum_{j=1}^{6} \; \sum_{k=1}^{60} \; \sum_{u=1}^{3} \; (in_{u} \cdot C_{i,j})=20$ \euro /day.
Therefore, it results that the company relocation cost is reduced by 23.5\%. The results of the perfomed simulations are shown in Table \ref{Inc}.

\begin{table}
\centering
\caption{ILP1 and ILP2 simulation results in Scenario 1}
\scalebox{1.1}{
\begin{tabular}{c|cc}
\hline 
& Relocation by staff & Relocation by users \\
& [\euro /day] & [\euro /day] \\
\hline 
Relocation Cost &  &  \\
for the Company & 34 & 26 \\  
\hline
Incentive for &  &  \\
the users & - & 20 \\  
\hline 
\end{tabular}} 
\label{Inc}
\end{table}

\begin{figure}[thpb]
      \centering
      \includegraphics[scale=0.35]{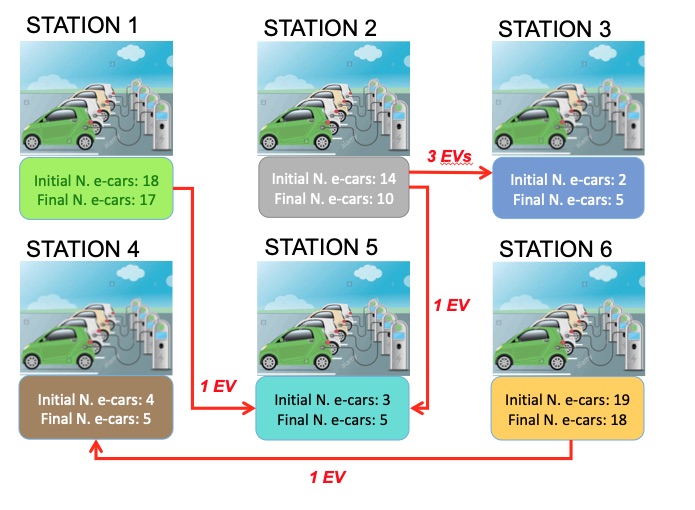}
      \caption{Electric car relocation by company staff (Scenario 1).}
      \label{EV_staff}
\end{figure}

\begin{figure}[thpb]
      \centering
      \includegraphics[scale=0.35]{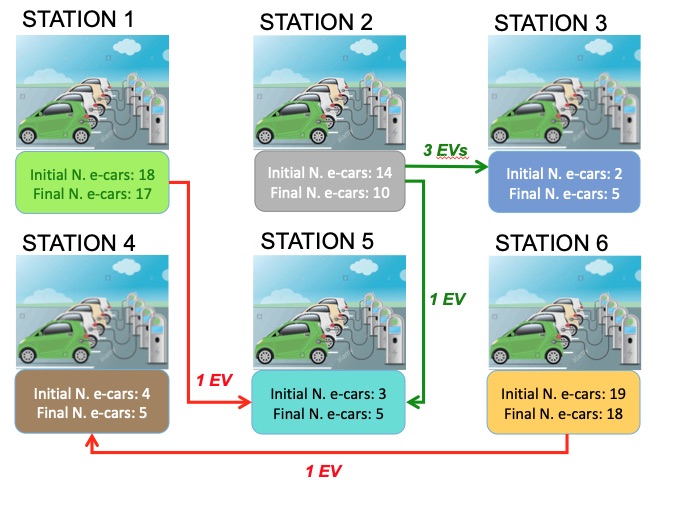}
      \caption{Electric car relocation process with user involvement (Scenario 1).}
      \label{EV_user}
\end{figure}

It is remarked that the performed EVs relocation trips are the same in both ILP1 and ILP2. The electric cars relocation solution is drawn in Figs \ref{EV_staff} and \ref{EV_user}, for ILP1 and ILP2, respectively. In particular, the red lines show the relocation tasks performed by the company staff while the green lines represents the relocation tasks performed by users. 

\subsection{SCENARIO 2: Electric Bike Sharing System}
In the second scenario, electric bikes must be relocated. By solving ILP1, the relocation cost for the sharing company by only using staff is $RC=25$ \euro /day. On the contrary, by involving users in the relocation process and solving ILP2 the relocation cost is $RCI=17.4$ \euro /day, with a corresponding reward $\sum_{i=1}^{6} \; \sum_{j=1}^{6} \; \sum_{k=1}^{60} \; \sum_{u=1}^{3} \; (in_{u} \cdot C_{i,j})=11.4$ \euro /day.

Hence, it results that the company relocation cost is reduced by 30.4\%.
The results of the simulations are shown in Table \ref{Scen2}.

\begin{table}
\centering
\caption{ILP1 and ILP2 simulation results in Scenario 2}
\scalebox{1.1}{
\begin{tabular}{c|cc}
\hline 
& Relocation by staff & Relocation by users \\
& [\euro /day] & [\euro /day] \\  
\hline 
Relocation Cost &  &  \\ 
for the Company & 25 & 17.4 \\
\hline 
Incentive for & &  \\ 
the users & - & 11.4 \\ 
\hline  
\end{tabular}} 
\label{Scen2}
\end{table}

\begin{figure}[thpb]
      \centering
      \includegraphics[scale=0.35]{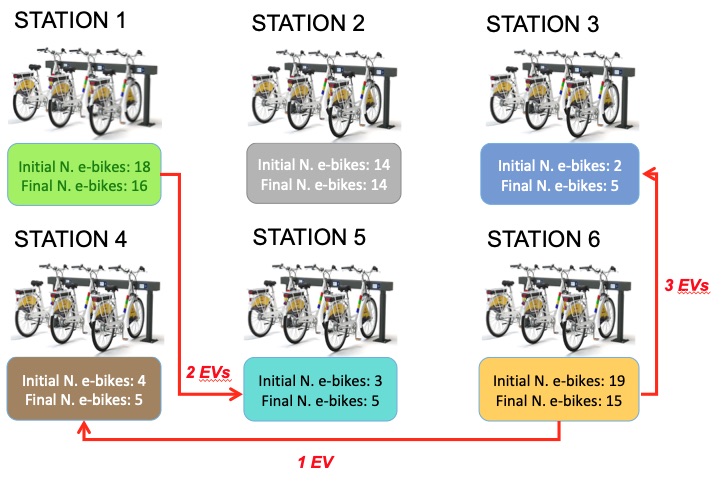}
      \caption{Electric car relocation by company staff (Scenario 2).}
      \label{ILP1_S2}
\end{figure}

\begin{figure}[thpb]
      \centering
      \includegraphics[scale=0.35]{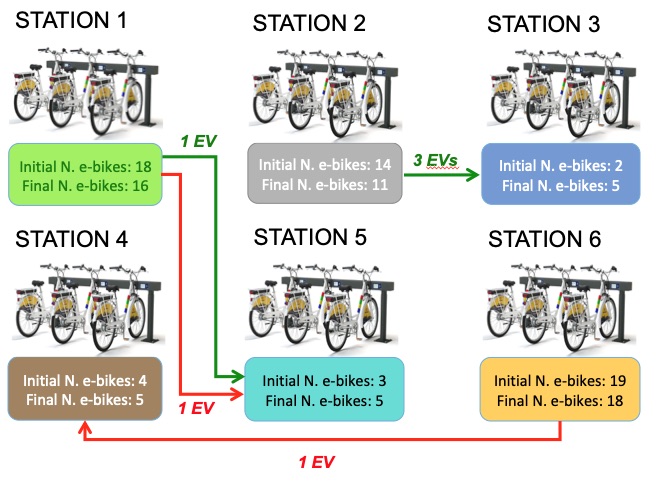}
      \caption{Electric car relocation process with user involvement (Scenario 2).}
      \label{ILP2_S2}
\end{figure}

The relocation tasks performed by solving ILP1 and ILP2 are respectively depicted in Fig. \ref{ILP1_S2} and Fig. \ref{ILP2_S2}. 
 
\subsection{Main Results Discussion}
It has been demonstrated that the proposed optimization approach to solve the EVRP for a mobility sharing company is able to reduce the EVs relocation cost for the company. This result has been achieved by transforming customers of the sharing service, from consumer to prosumer, incentivizing their participation in the relocation process. In this context, it is remarked that the application of the proposed incentive strategy is not a cost for the company and generate benefits for the users, in different forms e.g. credits to use the sharing service. 
It is noted that, despite the users participation in the relocation process, the company will maintain a high quality of service, by guaranteeing relocation performance by the company staff. 
In addition, the proposed approach allows for optimizing the relocation of EVs among the available stations.  
The results of the performed simulation in Scenario 1 and 2 show different relocation outcomes. This is due to the use of different kind of EVs involved in the two Scenarios. Indeed, the use small size EVs (Scenario 2) allows to pick-up the vehicles through vans by the company staff. This means that constraints (\ref{eq4}) of ILP1 and constraints (\ref{eq9}) of ILP2, indicating if an EV has enough battery autonomy to be moved from one station to another, can be relaxed in Scenario 2.

\section{Conclusion}
This paper presents an innovative electric vehicle (EV) relocation approach to minimize the relocation cost for the sharing company on the basis of user involvement by means of an incentive scheme. To this aim, two oInteger Linear Programming problems are formulated (ILP1 and ILP2). More in detail, the ILP1 allows to model and solve the EVRP by the company staff while the ILP2 upgrades ILP1 by adding the customers in the relocation process through an incentive mechanism. In this context, the users of the sharing service can travel for free and receive rewards if they accept to relocate EVs. Therefore, user availability will allow the sharing company to perform the EV relocation at a lower cost and will generate benefits for the user itself. In case of user unavailability the sharing service quality level will not decrease because the company will perform the vehicle relocation by the staff. 
A case study demonstrates that the relocation cost for the company is reduced if users are incentivized to participate in the relocation process of different kind of EVs. 

In future works, entry and exit rates of EVs will be included in order to improve the quality level of the sharing service for the final users  minimizing the company relocation cost, at the same time. 
Moreover, the optimization approach will take into account the free floating strategy.

\vspace{12pt}


\begin{thebibliography}{00}
\bibitem{Clemente} 
M. Clemente, M. P. Fanti, G. Iacobellis, M. Nolich and W. Ukovich, "A Decision Support System for User-Based Vehicle Relocation in Car Sharing Systems," in IEEE Transactions on Systems, Man, and Cybernetics: Systems, vol. 48, no. 8, pp. 1283-1296, Aug. 2018.
\bibitem{DSS1}
A. G. H. Kek, R. L. Cheu, Q. Meng, C. H. Fung, "A decision support system for vehicle relocation operations in carsharing systems,"
Transportation Research Part E: Logistics and Transportation Review, vol. 45, Issue 1, 2009, pp. 149-158.
\bibitem{two-way}
M. Nourinejad and M. J. Roorda, “Carsharing operations policies: A comparison between one-way and two-way systems,” Transportation, vol. 42, no. 3, pp. 497–518, 2015.
\bibitem{DSS2}
M. Nourinejad, M. J. Roorda, "A dynamic carsharing decision support system,"
Transportation Research Part E: Logistics and Transportation Review, vol. 66, 2014, pp. 36-50.
\bibitem{FF1}
S. Weikl, K. Bogenberger,
"A practice-ready relocation model for free-floating carsharing systems with electric vehicles – Mesoscopic approach and field trial results," Transportation Research Part C: Emerging Technologies, vol. 57, 2015, pp. 206-223.
\bibitem{CPLEX}
R. Zakaria, M. Dib, L. Moalic and A. Caminada, "Car relocation for carsharing service: Comparison of CPLEX and greedy search," 2014 IEEE Symposium on Computational Intelligence in Vehicles and Transportation Systems (CIVTS), Orlando, FL, 2014, pp. 51-58.
\bibitem{one_way1}
B. Boyacı, K. G. Zografos, N. Geroliminis,
"An optimization framework for the development of efficient one-way car-sharing systems,"
European Journal of Operational Research, vol. 240, Issue 3, 2015, pp. 718-733.
\bibitem{Dynamic_pr}
A. G. Bianchessi, S. Formentin and S. M. Savaresi, "Active fleet balancing in vehicle sharing systems via Feedback Dynamic Pricing," 16th International IEEE Conference on Intelligent Transportation Systems (ITSC 2013), The Hague, 2013, pp. 1619-1624.
\bibitem{Opt_Sim}
D. Jorge, G. H. A. Correia and C. Barnhart, "Comparing Optimal Relocation Operations With Simulated Relocation Policies in One-Way Carsharing Systems," in IEEE Transactions on Intelligent Transportation Systems, vol. 15, no. 4, pp. 1667-1675, Aug. 2014.
\bibitem{FF2}
F. Schulte, S. Voß, "Decision Support for Environmental-friendly Vehicle Relocations in Free-Floating Car Sharing Systems: The Case of Car2go," Procedia CIRP, vol. 30, 2015, pp 275-280.
\bibitem{one-way2}
A. Di Febbraro, N. Sacco and M. Saeednia, "One-Way Car-Sharing Profit Maximization by Means of User-Based Vehicle Relocation," in IEEE Transactions on Intelligent Transportation Systems, vol. 20, no. 2, pp. 628-641, Feb. 2019.
\bibitem{Incent}
A. Angelopoulos, D. Gavalas, C. Konstantopoulos, D. Kypriadis and G. Pantziou, "Incentivization schemes for vehicle allocation in one-way vehicle sharing systems," 2016 IEEE International Smart Cities Conference (ISC2), Trento, 2016, pp. 1-7.
\bibitem{one-way3}
Y. Deng, M.-A. Cardin, "Integrating operational decisions into the planning of one-way vehicle-sharing systems under uncertainty,"
Transportation Research Part C: Emerging Technologies, vol 86, 2018, pp. 407-424.
\bibitem{one-way4}
M. Xu, Q. Meng, Z. Liu, "Electric vehicle fleet size and trip pricing for one-way carsharing services considering vehicle relocation and personnel assignment," Transportation Research Part B: Methodological, vol. 111, 2018, pp. 60-82.
\bibitem{one-way5}
G. Homem de Almeida Correia, A. Pais Antunes,
"Optimization approach to depot location and trip selection in one-way carsharing systems,"
Transportation Research Part E: Logistics and Transportation Review, vol. 48, Issue 1, 2012, pp. 233-247.
\bibitem{Stack}
C. Boldrini, R. Incaini and R. Bruno, "Relocation in car sharing systems with shared stackable vehicles: Modelling challenges and outlook," 2017 IEEE 20th International Conference on Intelligent Transportation Systems (ITSC), Yokohama, 2017, pp. 1-8.
\bibitem{one-way6}
M. Bruglieri, A. Colorni, A. Luè,
"The Vehicle Relocation Problem for the One-way Electric Vehicle Sharing: An Application to the Milan Case," Procedia - Social and Behavioral Sciences, vol. 111, 2014, pp. 18-27.
\bibitem{one-way7}
M. Nourinejad, S. Zhu, S. Bahrami, M. J. Roorda,
"Vehicle relocation and staff rebalancing in one-way carsharing systems," Transportation Research Part E: Logistics and Transportation Review, vol. 81, 2015, pp. 98-113.
\bibitem{FF3}
A. G. Santos, P. G. L. Cândido, A. F. Balardino and W. Herbawi, "Vehicle relocation problem in free floating carsharing using multiple shuttles," 2017 IEEE Congress on Evolutionary Computation (CEC), San Sebastian, 2017, pp.2544-2551.
\bibitem{EU Reg}
Regulation (EU) N. 168/2013 of 15 January 2013 of the European Parliament and of the Concil url: https://eur-lex.europa.eu/LexUriServ/LexUriServ.do?uri=OJ:L:2013:060:0052:0128:
EN:PDF.
\end{thebibliography}
\end{document}